\documentclass[aps,prl,twocolumn,showpacs,groupedaddress]{revtex4}

\usepackage{graphicx}

\begin{document}

\preprint{12}

\title{Photon-induced vanishing of magnetoconductance in 2D electrons on liquid He}

\author{Denis Konstantinov}
\email[E-mail: ]{konstantinov@riken.jp}
\author{Kimitoshi Kono}
\affiliation{Low Temperature Physics Laboratory, RIKEN, Hirosawa 2-1, Wako 351-0198, Japan}

\date{\today}

\begin{abstract}
We report on a novel transport phenomenon realized by optical pumping in surface state electrons on helium subjected to perpendicular magnetic fields. The electron dynamics is governed by the photon-induced excitation and scattering-mediated transitions between electric subbands. In a range of magnetic fields, we observe vanishing longitudinal conductivity, $\sigma_{xx}\rightarrow 0$. Our result suggests the existence of radiation-induced zero-resistance states in the nondegenerate 2D electron system.    
\end{abstract}


\maketitle
\indent Electrons exhibit unique transport phenomena when they are confined in two dimensions and subjected to a strong perpendicular magnetic field. In a degenerate 2D electron gas (2DEG), the integer quantum Hall effect is characterized by exponentially small longitudinal resistivity, $\rho_{xx}\rightarrow 0$, and quantized Hall resistivity $\rho_{xy}$~\cite{PrangeGirvin}. In such a system, the electron transport is determined by the Fermi statistics of the charged carriers and the Landau quantization of their energy spectrum. In the quantum Hall regime, the suppression of scattering also results in vanishing diagonal conductivity, $\sigma_{xx}=\rho_{xx}/(\rho_{xx}^2+\rho_{xy}^2)$, and an electrical current flowing normal to the applied electric field~\cite{Tsui1982}. Recently, exponentially small photoinduced resistance and conductance of a degenerate 2DEG were discovered in ultrahigh-mobility GaAs/AlGaAs heterostructures \cite{Mani2002,Zudov2003,Yang2003} causing a surge of theoretical interest in this novel phenomenon~\cite{Andreev2003,Durst2003,Dorozhkin2003,Chepelianskii2009}. Here we report the occurrence of vanishing $\sigma_{xx}$ realized by inter-subband excitation in a system of nondegenerate electrons on liquid helium.
\newline
\indent Electrons on helium provide a unique classical counterpart to quantum Hall systems \cite{Andrei,MonarkhaKono}. The impurity-free environment results in an extremely high electron mobility, which, for sufficiently low temperatures, is limited only by the scattering of electrons from the quantized surface vibrations (ripplons) and exceeds $10^8$~cm$^2$V$^{-1}$s$^{-1}$. Unlike in semiconductors, electrons on helium retain their free-particle mass and g-factor. For a bulk helium substrate, the instability of the charged surface restricts the areal density of electrons to about $2\times 10^9$~cm$^{-2}$. The free-electron mass and low densities result in a very low Fermi energy, and at $T=0$, the system of interacting electrons favors the classical Wigner solid over the quantum degenerate regime~\cite{Williams1971}. This makes the observation of the quantum Hall effect impossible.
\newline 
\begin{figure}[b]
\centering
\includegraphics[width=8.0cm]{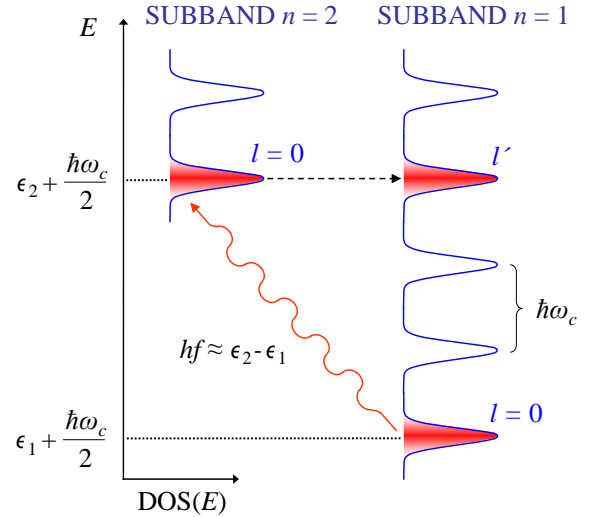}
\caption{\label{fig:1}(color online). Electron dynamics in perpendicular magnetic fields. The periodicity of the DOS reflects the single-electron energy spectrum given by Eq.~\ref{eq:1}. Microwave photons of energy $hf$ drive the inter-subband transition $n=1\rightarrow 2$ (wavy arrow) without changing the quantum state $l$. Excited electrons can be scattered elastically (dashed arrow) and fill the states $l'>l$ of the first subband.} 
\end{figure}
\indent Surface states of electrons are formed owing to the classical image potential, the repulsive barrier that prevents penetration inside the liquid, and the electric field $E_{\perp}$ applied perpendicular to the surface. In the resulting confinement potential, the electron dynamics is quantized into discrete electric subbands with energies $\epsilon_n$ $(n=1,2,...)$. At $E_{\perp}=0$, the energy spectrum is similar to that of the hydrogen atom, $\epsilon_n=-R/n^2$, where the effective Rydberg constant $R$ is approximately 7.6 and 4.2~K for liquid $^4$He and $^3$He, respectively. Below 1~K, almost all electrons are frozen into the lowest subband, and the inter-subband transition $1\rightarrow n$ can be excited using millimeter-wave microwaves with frequency $f=f_{n1}$, where $f_{n1}=(\epsilon_n-\epsilon_1)/h$ ($h$ is Planck's constant). In a magnetic field $B$ applied normal to the surface, the energy $E$ of an electron is affected by Landau quantization and the Zeeman effect and is given by
\begin{equation}
E=\epsilon_n+\hbar\omega_c(l+1/2)\pm g\mu_B B/2, \quad l=0,1,..~,
\label{eq:1}
\end{equation}
\noindent where $\omega_c=eB/m$ is the cyclotron frequency, $g\approx 2.0023$ is the g-factor, and $\mu_B=e\hbar/2m$ is Bohr's magneton. The Zeeman splitting $g\mu_B B$ is almost equal to $\hbar\omega_c$, therefore the density of states (DOS) possible for an electron occupying a subband of index $n$ consists of a sequence of collision-broadened peaks located at $\hbar\omega_c(l+1/2)$, as shown schematically in Fig.~\ref{fig:1}. The resonant absorption of microwave photons $f=f_{21}$ excites electrons into the states of the second subband, from which they undergo transitions back to the first subband mostly as a result of either the stimulated emission or scattering. The scattering processes are predominantly quasi-elastic, therefore the excited electrons are scattered into states having nearly the same energy as the initial states. This causes the filling of the high-index Landau levels of the $n=1$ subband  and alters the transport properties of the electron system. Changing $B$, the periodic variation of the inter-subband scattering rate, which accompanies the sequential alignment of the Landau levels of the two subbands, results in the conductance oscillations recently reported by the authors \cite{KonstantinovOSC2009}. The conductivity $\sigma_{xx}$ of the irradiated electrons was found to vary periodically with the ratio $2\pi f/\omega_c$ and showed a strong dependence on the temperature $T$ and electron density $n_s$. Increasing either $T$ or $n_s$ led to the disappearance of the oscillations owing to collision broadening of the Landau levels or many-electron effects.
\newline
\indent In this Letter, we report the observation of a novel transport effect characterized by vanishing $\sigma_{xx}$ in electrons on liquid $^3$He cooled to below 0.3~K. In the certain intervals of $B$, the magnetoconductance of irradiated electrons rapidly decreases with decreasing $T$ and increasing radiation intensity until $\sigma_{xx}$ abruptly drops to zero and exhibit a hysteresis in varying magnetic fields.
\newline
\indent The diagonal conductivity $\sigma_{xx}$ of electrons is measured using the Sommer-Tanner technique \cite{SommerTanner} adapted for the Corbino geometry. A circular pool of electrons is formed on the free surface of liquid $^3$He held midway between two circular parallel plates, each having a diameter of 20~mm, separated by $d=2.6$~mm. The Corbino disk forms the top plate and consists of two concentric electrodes separated by a gap 0.2~mm wide. An ac (0.1-1~kHz) voltage $V_{in}$ of 10~mV rms is applied to one electrode, inducing an ac current $I_{out}$, which is typically on the order of 1~pA, that flows to the other electrode through the sheet of electrons. The components of $I_{out}$ at the phase angles of $0^\circ$ and $90^\circ$ relative to $V_{in}$ are measured using a lock-in amplifier. For a perfectly conducting electron sheet, the coupling between electrodes is purely capacitive and the in-phase component is zero. For a finite $\sigma_{xx}$, a resistive component of $I_{out}$ appears at a phase angle of $0^\circ$. The relationship between the complex admittance $G=I_{out}/V_{in}$ and $\sigma_{xx}$ is determined by assuming azimuthal symmetry and solving the electrodynamic problem of electric field distribution inside the experimental cell \cite{Mehrotra1987}. The experiment is carried out in  magnetic fields of up to 0.85~T produced by a superconducting coil placed around the cell. The value of $B$ is obtained from the coil current, $I$, via the calibration constant $\kappa=B/I$. The latter is determined accurately by {\it in situ} measurements of the cyclotron resonance (CR) of electrons on helium. For this purpose, an rf signal (10-20~GHz) from the synthesized signal generator is applied to the bottom electrode, and the CR-induced change in the conductivity signal is recorded by sweeping the magnetic field~\cite{Penning1998}.
\newline
\begin{figure}[t]
\centering
\includegraphics[width=8.0cm]{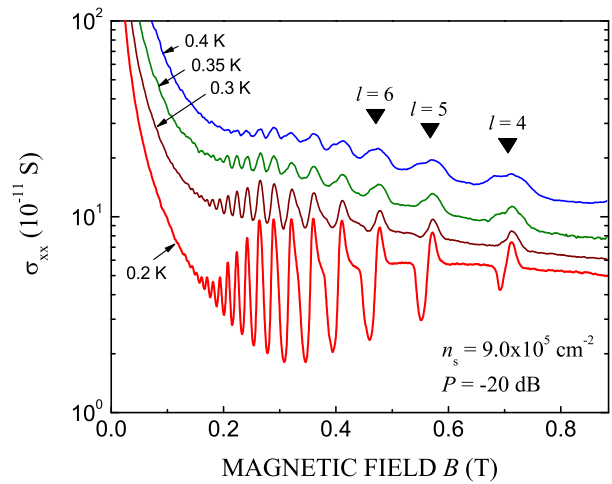}
\caption{\label{fig:2} (color online) Longitudinal conductivity $\sigma_{xx}$ versus $B$ for irradiated electrons with $n_s=0.9\times 10^6$~cm$^{-2}$ at four different temperatures: $T=0.4$ (blue line), 0.35 (green line), 0.3 (brown line), and 0.2~K (red line). Black triangles indicate the values of $B$ where $hf=l\hbar \omega_c$, for $l=4$, 5, and 6.} 
\end{figure}
\indent Electrons are generated by thermionic emission from a filament placed above the liquid. The bottom plate is biased at a positive voltage $V$, and the electron density $n_s$ is determined from the shielding condition of the electric field above the surface, $n_s=\varepsilon V/2\pi ed$, where $\varepsilon$ is the dielectric constant of the liquid. After the surface is charged, the transition frequency of electrons $f_{21}$ is tuned to the resonance with a microwave frequency $f$ of 79~GHz by adjusting the voltage $V$. Microwaves are transmitted from the source (with a maximum output power of about 5~dBm) along a waveguide into the cell, yielding a maximum power of about -10~dBm at the sample.
\newline
\indent As was previously reported~\cite{KonstantinovOSC2009}, at $T\geq 0.5$~K, where the DOS functions of two subbands significantly overlap owing to collision broadening of the Landau levels, the oscillatory part of $\sigma_{xx}$ follows a sequence of maxima (minima) as the frequency ratio $2\pi f/\omega_c$ attains successive integer (half-integer) values. This behavior reflects the periodic increase in the inter-subband scattering of the microwave-excited electrons as the energy levels of two subbands undergo sequential alignment. In addition, we showed that corrections to the single-electron energy coming from the many-electron fluctuating electric field strongly affect the oscillations at high electron densities.
\newline
\indent Figure~\ref{fig:2} shows $\sigma_{xx}$ versus $B$ for electrons with $n_s=0.9\times 10^6$~cm$^{-2}$ for radiation power $P$ of -20~dB, which is measured at the microwave source and expressed as a ratio of the maximum power, at several temperatures below 0.5~K. For such a low $n_s$ and moderate $B$, the many-electron effects become relatively unimportant, and the shape of the oscillations is determined by the level broadening due to electron scattering. The latter decreases rapidly with cooling of the liquid as the concentration of scattering particles (helium vapor atoms and ripplons) decreases. Correspondingly, upon decreasing the temperature from 0.4 to 0.2~K, the oscillations develop into a sequence of narrow maxima located near the fields satisfying the commensurability condition for the energy $hf=l\hbar\omega_c$ (cf. Fig.~\ref{fig:1}). For $T=0.4$, 0.35, and 0.3~K, the total scattering rates at zero field $\nu_0$ are 4.0, 1.6, and 0.6$\times 10^8$~s$^{-1}$, respectively. Correspondingly, assuming the single-electron approximation and short-range scattering from vapor atoms, the width of the Landau level $\Gamma=\hbar (2\omega_c\nu_0/\pi)^{1/2}$ decreases by a factor of 2.6, which is in qualitative agreement with Fig.~\ref{fig:2}. At $T=0.2$~K, the scattering is predominantly due to ripplons, and the scattering rate is about $0.2\times 10^8$~s$^{-1}$. In this regime, the width $\Gamma$ depends on the Landau index $l$ \cite{MonarkhaKono}. For $l=4$, we estimate that $\Gamma\approx 0.015$~T, which also roughly agrees with the width of the corresponding peak in Fig.~\ref{fig:2}. 
\begin{figure}[b]
\centering
\includegraphics[width=8.0cm]{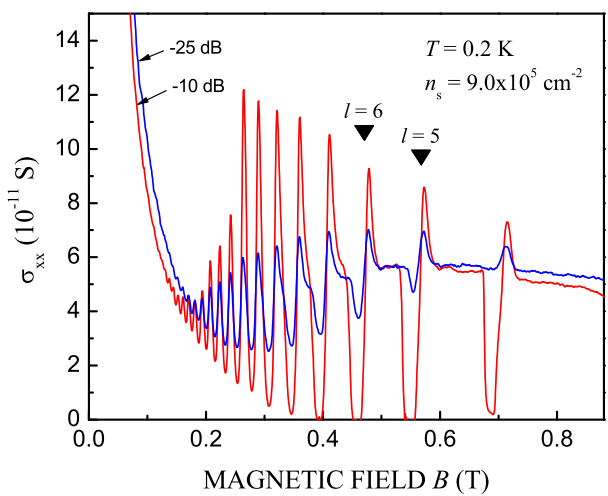}
\caption{\label{fig:3} (color online) $\sigma_{xx}$ versus $B$ for $n_s=0.9\times 10^6$~cm$^{-2}$, $T=0.2$~K, and radiation power $P=-25$ (blue line) and -10 dB (red line).}
\end{figure}
\begin{figure}[b]
\centering
\includegraphics[width=8.0cm]{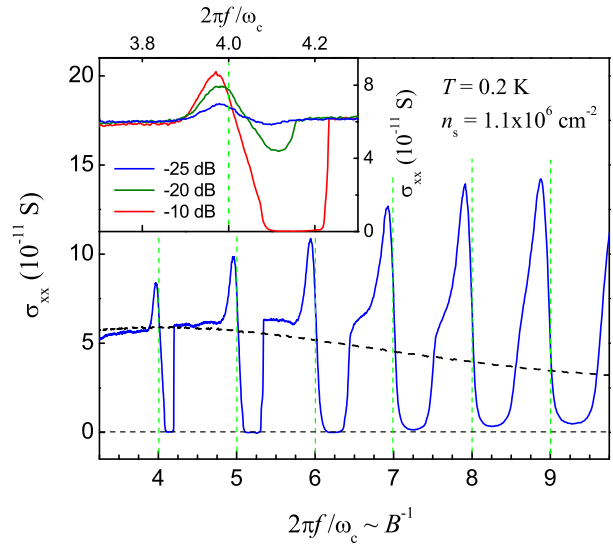}
\caption{\label{fig:4} (color online) $\sigma_{xx}$ versus $2\pi f/\omega_c$ obtained without radiation (dashed line, black) and with radiation of frequency $f\approx f_{21}$ at $P= -10$~dB (solid line, blue). Inset: $\sigma_{xx}$ in the vicinity of $2\pi f/\omega_c=4$ for three different power levels.} 
\end{figure}
\newline
\indent At the fixed value of $P$, the shape of oscillations becomes more complicated with decreasing $T$. In addition to conductivity maxima corresponding to $hf=l\hbar\omega_c$, a marked decrease in $\sigma_{xx}$ was observed on the low-field side of each maximum. This behavior is illustrated in Fig.~\ref{fig:2} by the bottom curve obtained at $T=0.2$~K. At the resulting minima, $\sigma_{xx}$ decreases rapidly with increasing $P$. At $T=0.2$~K, this new effect produces giant oscillations of $\sigma_{xx}$, which, for sufficiently high powers, approaches zero in certain intervals of magnetic fields. Figure~\ref{fig:3} shows $\sigma_{xx}$ versus $B$ for $n_s=0.9\times 10^6$~cm$^{-2}$ and two different values of $P$. For $P=-10$~dB, the plot demonstrates vanishing $\sigma_{xx}$ on the low-field side of the maxima corresponding to $l=5$ and 6. We emphasize that neither oscillations nor zero-conductance states are observed when $f_{21}$ of the electrons is tuned away from $f$.
\newline
\indent As was previously shown~\cite{KonstantinovBIST2009}, in the Drude regime, the heating of electrons with resonant microwaves leads to the thermal population of higher excited subbands and significantly alters electron transport along the surface. At $T=0.2$~K, this can lead to a decrease in the scattering of electrons, and therefore a decrease in $\sigma_{xx}$, as electrons in the higher subbands localize farther from the surface and their interaction with ripplons weakens. In strong $B$, the quantization of the lateral motion of electrons significantly complicates analysis. It is possible that similar effects associated with electron heating arise under the conditions of the present experiment. However, further investigation is required to elucidate the role of heating in the formation of the conductance minima reported here.
\newline   
\indent Figure~\ref{fig:4} shows $\sigma_{xx}$ obtained for $n_s=1.1\times 10^6$~cm$^{-2}$ at $T=0.2$~K with and without radiation, plotted versus $2\pi f/\omega_c$. To obtain these plots, $f$ is centered at the inter-subband resonance, which has a full width at half maximum of about 0.3~GHz. $B$ was slowly swept at a rate of $2\times 10^{-5}$~Ts$^{-1}$ to ensure that the result is not affected by the time constant, $T=10$~s, of the lock-in amplifier. In the range of $2\pi f/\omega_c$ shown in Fig.~\ref{fig:4}, the oscillations exhibit periodicity in the inverse magnetic field $B^{-1}$. Unlike in the high-$T$ regime previously reported~\cite{KonstantinovOSC2009}, the maxima of $\sigma_{xx}$ appear to be shifted to the left of integral $2\pi f/\omega_c$ owing to the formation of deep minima on the right side of each peak. This behavior is illustrated in the inset of Fig.~\ref{fig:4}, where $\sigma_{xx}$ is plotted in the vicinity of $2\pi f/\omega_c=4$ for three different values of $P$.
\begin{figure}[t]
\centering
\includegraphics[width=8.0cm]{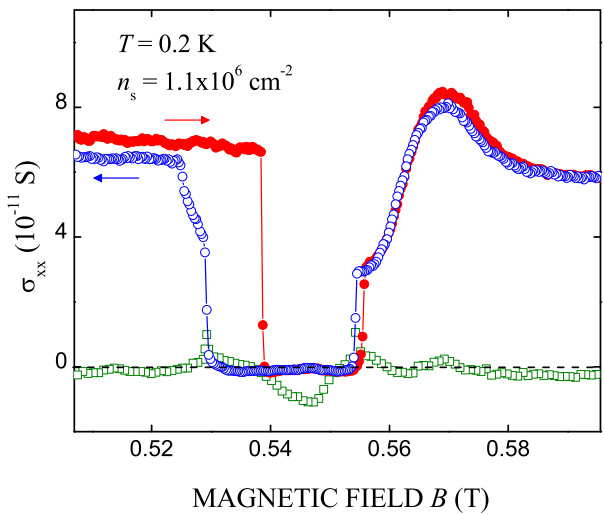}
\caption{\label{fig:5} (color online) $\sigma_{xx}$ obtained in an upward (solid circles, red) and downward (open circles, blue) sweeps of $B$ at $T=0.2$~K, $n_s=1.1\times 10^6$~cm$^{-2}$ and $P=-10$~dB. In zero-resistance regime, $\sigma_{xx}$ is complex with a negative imaginary part (squares) and a real part (circles) attaining slightly negative values (see explanation in the text).} 
\end{figure}
\newline
\indent In the zero-conductance regime, the measured in-phase component of $I_{out}$ drops to slightly negative values. Simultaneously, the quadrature component shows negative values. Provided that our assumption of azimuthal symmetry is correct, the data analysis indicates a complex conductivity $\sigma_{xx}$ with a negative imaginary part and a real part having a very small negative value. The real part of $\sigma_{xx}$ shows a strong dependence on the direction of the magnetic field sweep, as illustrated in Fig.~\ref{fig:5} for the minimum $l=5$. Here, the real part of $\sigma_{xx}$ is shown for the upward sweep (closed circles), while both real (solid circles) and imaginary (open squares) parts of $\sigma_{xx}$ are shown for the downward sweep. The measurements are carried out at a sweep rate of approximately $10^{-5}$~Ts$^{-1}$ to eliminate the effect of the time constant of the measurement system. Upon slowly increasing $B$, $\sigma_{xx}$ abruptly drops to below zero and vanishes, within the range of experimental uncertainty, in a certain interval of fields. Upon the downward sweep of $B$, $\sigma_{xx}$ retains the vanishing value down to significantly lower fields. This hysteretic behavior is reminiscent of correlation-induced optical bistability recently observed in electrons on helium~\cite{KonstantinovBIST2009}. This indicates that the heating of electrons and many-electron effects might be important in the formation of zero-conductance states.  
\newline
\indent The linear off-diagonal conductivity, $\sigma_{xy}=n_se/B$, has been confirmed in electrons on helium in the Hall bar setup~\cite{Lea1988}. For $n_s=0.9\times10^6$~cm$^{-2}$ and $B=0.85$~T, this gives $\sigma_{xy}\approx 10^{-9}$~S. Although we were unable to measure $\sigma_{xy}$ in the present geometry, it seems reasonable to assume that $\sigma_{xy}$ remains finite under the conditions of our experiment. Then, the vanishing $\sigma_{xx}$ also implies a vanishing diagonal resistivity $\rho_{xx}$~\cite{Yang2003}. Therefore, the effect reported here may be related to radiation-induced zero-reistance states found in the degenerate 2DEG in semiconductors~\cite{Mani2002,Zudov2003,Andreev2003,Durst2003,Dorozhkin2003,Chepelianskii2009}. 
\newline 
\indent In summary, we observed the vanishing of the diagonal conductivity, $\sigma_{xx}\rightarrow 0$, in a system of nondegenerate electrons on liquid helium. The effect is induced by the inter-subband absorption of microwaves and appears in the ranges of $B$ where the energy difference between two subbands, $hf_{21}$, exceeds the integral cyclotron energy, $l\hbar\omega_c$. The vanishing conductance appears at low $T$ and high intensities and exhibits hysteresis in varying $B$.
\newline
\indent We acknowledge discussions with Yu.~P. Monarkha and M.~I. Dykman. This work was supported in part by MEXT through Grants-in-Aid for Scientific Research.

%\bibliography{refdata}

\begin{references}

\bibitem{PrangeGirvin} {\it The Quantum Hall Effect}, edited by R.~E. Prang and S.~M. Girvin (Springer, New York, 1990).

\bibitem{Tsui1982} D.~C. Tsui, H.~L. Stormer, and A.~C. Gossard, Phys. Rev. B \textbf{25}, 1405 (1982).

\bibitem{Mani2002} R.~G. Mani {\it et al.}, Nature (London) \textbf{420}, 646 (2002).

\bibitem{Zudov2003} M.~A. Zudov, R.~R. Du, L.~N. Pfeifer, and K.~W. West, Phys. Rev. Lett. \textbf{90}, 046807 (2003).

\bibitem{Yang2003} C.~L. Yang {\it et al.}, Phys. Rev. Lett. \textbf{91}, 096803 (2003).

\bibitem{Andreev2003} A.~V. Andreev, I.~L. Aleiner, and A.~J. Millis, Phys. Rev. Lett. \textbf{91}, 056803 (2003).

\bibitem{Durst2003} V.~I. Ryzhii, Sov. Phys. Solid State \textbf{11}, 2078 (1970); A.~C. Durst, S. Sachdev, N. Read, and S.~M. Girvin, Phys. Rev. Lett. \textbf{91}, 086803 (2003); X.~L. Lei and S.~Y. Liu, {\it ibid} \textbf{91}, 226805 (2003); M.~G. Vavilov and I.~L. Aleiner, Phys. Rev. B \textbf{69}, 035303 (2004). 

\bibitem{Dorozhkin2003} S.~I. Dorozhkin, JETP Lett. \textbf{77}, 681 (2003); I.~A. Dmitriev {\it et al.}, Phys. Rev. B \textbf{71}, 115316 (2005).

\bibitem{Chepelianskii2009} A.~D. Chepelianskii and D.~L. Shepelyansky, Phys. Rev. B \textbf{80}, 241308(R) (2009). 

\bibitem{Andrei} {\it Electrons on Helium and Other Cryogenic Substrates}, edited by E.~Y. Andrei (Kluwer Academic, Dordrecht, 1997).

\bibitem{MonarkhaKono} Yu. P. Monarkha and K. Kono, {\it Two-Dimensional Coulomb Liquids and Solids} (Springer, Berlin, 2004).

\bibitem{Williams1971} R. Williams and R.~S. Crandall, Phys. Lett. A \textbf{36}, 35 (1971); also, see review by V. Shikin in \cite{Andrei}.

\bibitem{KonstantinovOSC2009} D. Konstantinov and K. Kono, Phys. Rev. Lett. \textbf{103}, 266808 (2009).

\bibitem{SommerTanner} W.~T. Sommer and D.~J. Tanner, Phys. Rev. Lett. \textbf{27}, 1345 (1971).

\bibitem{Mehrotra1987} R. Mehrotra and A.~J. Dahm, J. Low Temp. Phys. \textbf{67}, 641 (1987); L. Wilen and R. Giannetta, {\it ibid.} \textbf{72}, 353 (1988).

\bibitem{Penning1998} F.~C. Penning, O. Tress, H. Bluyssen, and P. Wyder, J. Low Temp. Phys. \textbf{110}, 185 (1998).

\bibitem{KonstantinovBIST2009} D. Konstantinov {\it et al.}, Phys. Rev. Lett. \textbf{103}, 096801 (2009).

\bibitem{Lea1988} M.~J. Lea, A.~O. Stone, and P. Fozooni, Europhys. Lett. \textbf{7}, 641 (1988); R.~W. van der Heijden, H.~M. Gijsman, and F.~M. Peeters, J. Phys. C: Solid State Phys. \textbf{21}, L1165 (1988).
 

\end{references}
%

\end{document}